# Nonreciprocal Transport with Quantum Geometric Origin in Layered Hybrid Perovskite

Zihan Zhang[1*], Sihan Chen[2*], Mingfeng Chen[3], Jee Yung Park[4,5], Gang Shi[1], Kaitai Xiao[1], Swati Chaudhary[6], Alejandro T. Busto[7], Kenji Watanabe[8], Takashi Taniguchi[9], Peng Xiong[1,10], Xiao-Xiao Zhang[7], Efstratios Manousakis[1], Letian Dou[11,12], Xi Wang[3,13], Cyprian Lewandowski[1,10,14], and Hanwei Gao[1,10]

1 Department of Physics, Florida State University, Tallahassee, FL 32306, USA.

2 Department of Physics, University of Chicago, Chicago, Illinois 60637, USA.

3 Department of Physics, Washington University, Saint Louis, Missouri 63130, USA.

4 Department of Chemical and Environmental Engineering, Yale University, New Haven, CT 06511, USA

5 Energy Sciences Institute, Yale University, West Haven, CT, USA

6 Institute for Solid State Physics, The University of Tokyo, Kashiwa, Chiba, 277-8581, Japan

7 Davidson Department of Physics, University of Florida, Gainesville, FL 32611, USA.

8 Research Center for Electronic and Optical Materials, National Institute for Materials Science, 1-1 Namiki, Tsukuba 305-0044, Japan.

9 Research Center for Materials Nanoarchitectonics, National Institute for Materials Science, 1-1 Namiki, Tsukuba 305-0044, Japan

10 FSU Quantum Initiative, Florida State University, Tallahassee, FL 32306, USA.

11 Department of Chemistry, Emory University, Atlanta, GA 32322, USA

12 Davidson School of Chemical Engineering, Purdue University, West Lafayette, IN 47907, USA.

13 Institute of Materials Science and Engineering, Washington University, St. Louis, Missouri 63130, USA

14 National High Magnetic Field Laboratory, Tallahassee, FL, 32310, U.S.A.

**Abstract**

Quantum geometry quantifies how the electron wavefunction evolves distinctly from conventional transport theory. In noncentrosymmetric materials, nonreciprocal transport with quantum geometric origin remains prominent with localized charge independent of vanished group velocity. The discovery of such nonreciprocal and nonlinear responses was realized by recent advances in two-dimensional materials. As a promising candidate, the electronic structure and symmetry of layered hybrid perovskites can be deliberately designed and manipulated by incorporating selected organic ligands. Despite the observation of exotic photogalvanic effects and chiral optical effects, the underlying mechanism how these nonlinear responses are enabled in the multi-quantum well structures remained unclear. Here we demonstrated the quantum geometric origin for interlayer spontaneous photocurrent in $(PEA)_2PbI_4$. Contrary to assumptions that charge transport across the 2D planes is limited, we observed a spontaneous photocurrent along this crystalline orientation. Theoretical analysis using a tight-binding model identifies shift current as the microscopic origin. This quantum geometric effect is enabled by ionic displacements from centrosymmetric coordinates and enhanced by multiband transition high-density bands of the layered hybrid crystal. We anticipate that such unique low-dimensional systems with structure can provide fertile ground for discovering novel optoelectronic functionalities.

**Main**

Quantum geometry is a vital concept in condensed matter physics and raised research interest in order to understand the emerging superfluid weight in flat bands[1,2] and quantum Hall effects[3,4]. It quantifies how electron wavefunction evolves upon change in parameters in a fundamentally different way from conventional transport theory. One example is the shift current which facilitates unidirectional charge transport under light excitation without an applied or built-in electric field in noncentrosymmetric materials.[5] The spontaneous photocurrent stems from the coordinate shift of photoexcited electrons due to difference in quantum geometric quantities during interband transitions.[6] Distinct from classical transport in dispersive bands, the quantum geometric origin of shift current ensures spontaneous photocurrent even without free carriers, such as with strong localization[7] and below-bandgap excitation[8]. Such nonreciprocal responses not only are macroscopic manifestations of quantum mechanical effects, but can also serve as a sensitive probe of new quantum phases of materials.[9]

Low-dimensional materials have been instrumental in studying nonreciprocal and nonlinear effects. Symmetry-breaking structures are more accessible by assembling the 2D building blocks artificially into staggered or twisted lattices.[10,11] More recently, quasi-2D halide perovskites offered new opportunities to manipulate the electronic structure and crystal symmetry in a self-assembled manner. These organic-inorganic hybrid materials belong to a new class of semiconductors that have drawn significant attention recently for their potential in sustainable energy applications.[12,13] These hybrid crystals consist of alternating layers of organic and inorganic sublattices (**Fig. 1a**), forming highly ordered multi-quantum wells.[14] Their crystal and electronic structures are sensitively affected the types of organic ligands incorporated, because of the soft ionic bonding framework and strong intermolecular interactions.[15] Pioneering works have demonstrated various photogalvanic and chiral optoelectronic effects by introducing polar or chiral cations into the layered perovskites.[16,17] The microscopic origin of the phenomena, however, remains elusive in these new types of low-dimensional crystals.

In this work, we uncovered the quantum geometric nature of nonreciprocal responses in the layered perovskite $(PEA)_2PbI_4$. Charge transport across the 2D planes was previously considered to be limited and largely disregarded, because of the minimal band dispersion. We observed, alternatively, spontaneous photocurrent along this crystalline orientation. The unidirectional transport was accompanied by optical second harmonic generation, both of which were manifestations of the second-order nonlinear character. Theoretical analysis based on a tight-binding model showed that shift current is the underlying mechanism of the observed photoresponse.[6] In the model, ions displaced from their centrosymmetric coordinates imparted a non-trivial quantum geometry of the electronic bands. Interestingly, the quantum geometry may also be carried by bound charges, reflected as the spontaneous photocurrent generated from sub-bandgap photons.[18,19] The shift current in the layered hybrid crystal, based on the model, is dominated multiband processes instead of the direct two-band transitions. This is enabled by its unique high-density and nondispersive bands and offers a new pathway for enhancing the shift current or related nonreciprocal responses.

**Device fabrication**

To characterize the cross-plane charge transport in $(PEA)_2PbI_4$, two-terminal devices were constructed with layered architecture (*Methods*). Thin flakes of the layered perovskite crystal were sandwiched between few-layer-graphene electrodes (**Fig. 1a**). The flakes were exfoliated from monolithic crystals of $(PEA)_2PbI_4$ and cleaved naturally along the 2D planes. The lateral dimensions were about 10 μm across (**Fig. 1b**). The thickness, determined using atomic force microscopy, was on the order of 100 nm (**Fig. 1c**). The devices were encapsulated using thin flakes of hexagonal boron nitride (hBN) to prevent the perovskite from degradation in air.[20]

The layered perovskite exhibited pronounced photoresponse when illuminated with 355-nm laser (**Fig. 2a**). While the resistance was greater than 100 TΩ in the dark, it dropped by more than three orders of magnitude under illumination, reaching 23 GΩ with an optical power of 50 μW. Such high dark resistance is consistent with the nearly flat bands along the out-of-plane axis of the layered crystal, as shown in the tight-binding calculation discussed. The large contrast between the dark and photoconductivity indicated high crystallinity and low density of trap states in the layered perovskite.

**Observation of nonreciprocal and nonlinear responses**

The photocurrent carried a nonreciprocal character. It came with an open-circuit voltage ($V_{oc}$) consistently at 50 mV and a short-circuit current ($I_{sc}$) that scaled linearly with the power of illumination (**Fig. 2a&b**). As a signature of the photovoltaic effect, $V_{oc}$ was commonly observed in semiconductor devices with p-n or Schottky junctions.[21] It was, however, not expected in our devices provided the symmetrical graphene-perovskite-graphene architecture. Note that the phenomenon was distinct from the anomalous photovoltaic effect induced by ion migration reported earlier.[22,23] The current-voltage hysteresis accompanying the ion migration was not observed in our measurements. With these possibilities precluded, we speculated that the photovoltage observed here originated from the inversion symmetry breaking inherent to the perovskite itself. What aligned with this hypothesis was the polarization dependence of the photocurrent. As the polarization angle of linearly polarized light was varied, the $I_{sc}$ exhibited a two-fold polar symmetry (**Fig. 2c**, symbols), indicating the presence of an effective crystal polarization.

The absence of inversion symmetry was also evident in optical second harmonic generation (SHG). SHG characterizes nonlinear optical responses at twice the frequency of excitation light $\omega$. The intensity of SHG can be expressed as[24]

$$I_{SHG} \propto \left|\epsilon_0 \chi_{ijk}^{(2)} E_j(\omega) E_k(\omega)\right|^2, \qquad \text{(Eqn. 1)}$$

where $E$ is the electric field of incident light. The nonlinear susceptibility tensor ($\chi_{ijk}^{(2)}$) is non-zero only if the system is lack of inversion symmetry, which makes SHG a sensitive probe of noncentrosymmetric crystals.[24] When (PEA)$_2$PbI$_4$ flakes were excited using a 784-nm pulsed laser, pronounced SHG signals were detected at 392 nm (**Fig. 2d**). The intensity scaled quadratically with the excitation power (**Fig. 2e**), consistent with the power dependence described by Eqn.1. When the polarization angle of excitation was varied, the SHG showed an anisotropic polar pattern similar to that of spontaneous photocurrent (**Fig. 2f**). It indicated that the same (in-plane) crystal polarization could be responsible for both the nonreciprocal charge transport and the nonlinear optical response.

**Construction of an effective tight-binding model**

To uncover the microscopic origin of the nonreciprocal transport, we modeled the layered perovskite using an effective tight-binding model.[25] (PEA)$_2$PbI$_4$ in previous works has been treated nominally as a centrosymmetric crystal, although distortions of the corner-sharing octahedra were speculated in some reports.[26] We confirmed using the tight-binding model that, if the inversion symmetry was preserved, both the spontaneous photocurrent and SHG responses would vanish (*Supplementary note 4* and **Fig. S1**).

In our model, ions were displaced from their centrosymmetric positions to break the inversion symmetry. Specifically, the lead ions (Pb$^{2+}$) were shifted from the octahedral center by an in-plane vector of ($\delta_x$, $\delta_y$) (**Fig. 3a**); the top and bottom iodine ions (I$^-$) were sheared by $\delta_s$ along the diagonal orientation of the equatorial plane (**Fig. 3b**). Such lattice distortion resembled the tilt of [PbI$_6$]$^{4-}$ octahedra reported previously.[27] In addition, the organic ligands gave rise to a staggered structure between the neighboring inorganic layers. These structural deformations together broke the out-of-plane mirror symmetry and resulted in the unit cell of the Bravais lattice twice as large along the z axis (Fig. 3b). While this pattern of structural deformations was phenomenological, the physical behaviors calculated accordingly reproduced the experimental observations quite well.

The tight-binding model was first validated by the comparison with results from density functional theory in our previous work.[25] In the case of centrosymmetric lattices, the band dispersion calculated using the two methods agreed well with each other. When the ions were displaced, the valence bands were significantly modified, while the low-energy conduction bands remained nearly unchanged (**Fig. 3c**). The bands close to the charge neutrality (the edges of bandgap) were dominated by the hybridized orbitals between I (5p) and Pb (6s).[28] These bands dictate electrical conductivity in semiconductors. Minimal contribution from the organic constituents to these low-energy states allowed us to simplify the calculation by neglecting the $PEA^+$ ligands in our tight-binding model. An effective coupling ($I_z$) was introduced between the adjacent inorganic layers. Such coupling was necessary to facilitate out-of-plane charge transport between the layers, which would otherwise be electronically decoupled according to the calculated band dispersion.

**Calculation of shift vector and shift conductivity**

In noncentrosymmetric crystals, injection current and shift current are two primary intrinsic mechanisms behind spontaneous photocurrent. The former requires illumination from circularly polarized light.[5] The spontaneous photocurrent we observed here, alternatively, sustained withe reversal symmetry preserved, agreeing better with the behavior of shift current. Other mechanisms for the bulk photovoltaic effect, related to momentum transfer or asymmetric distribution of momentum for examples, are less likely to be responsible either. These extrinsic origins rely on diffusive charge transport, which expect to be suppressed with the flat electronic bands calculated between the Γ and Z points (Fig. 3b) and the minimized electron velocity.

As a type of second-order optoelectrical response, shift current differs fundamentally from the conventional diffusive charge transport. The magnitude of shift current can be quantified by calculating the so-called shift vector using our model (essentially the integrand of the shift current expression, *Supplementary note 1*).[29] When the shift vector is finite, electrons experience real-space shift upon optically excited interband transitions,[6] giving rise to a net electric current without applied or built-in electric fields.[5,30] It is a manifestation of the quantum geometry,[31] which enables collective motion of charge carriers even if they appeared to be localized under the traditional band transport framework.[32]

The shift vector, calculated with the $Pb^{2+}$ and $I^-$ ions displaced, showed asymmetric distributions in the reciprocal space (**Fig. 3d**). Here the shift vector was plotted by taking an average of $\sigma_{zxx}$ and $\sigma_{zyy}$ in the shift conductivity tensor to account for the shift current response under unpolarized light (*Supplementary note 2*). Around the Γ point, contributions were dominated along diagonal orientations where the impact of inversion symmetry breaking was maximized. The distribution also exhibited strong dependence on the energy of interband transitions ($E$). Near the optical bandgap ($E - E_g = 0.005$ eV), finite shift vectors were more localized near the Γ point. At higher energy, greater shift vectors were found with larger momentum away from Γ. This is consistent with the calculated band dispersion, in which the conduction band minima and valence band maxima are located at the Γ point. Note that the magnitude of shift vectors increased with $E$, indicating that the strongest shift current would not occur near the bandgap, but at a higher energy. This feature is more explicit in the spectra of calculated and measured shift current.

**Comparison between theoretical and experimental results**

The theoretically calculated shift conductivity agreed well with the measured photocurrent spectrum (**Fig. 3e**). The total shift conductivity was calculated by taking an integral of the shift vector across the Brillouin zone. It was calculated as a function of the ion displacements ($δ_x$, $δ_y$, $δ_s$) and the energy of interband transitions ($E$). Note that the calculated bandgap energy is smaller than the measured, which is a known limitation of the density functional theory.[33] The comparison between the calculated and measured spectra provided the basis for optimizing the ion displacements in the tight-binding model (*Supplementary note 3*). The best agreement was found with $δ_x = 0.01a$, $δ_y = 0.02a$, $δ_s = 0.03a$ and $I_z = -2.11$ eV. $a$ is the in-plane lattice constant of the unit cell containing four tilted $[PbI_6]^{4-}$ octahedra (Fig. 3a), and is 8.74 Å according to crystallographic measurements reported previously.[27] Such small displacements could be challenging to be resolved using diffractometry techniques, especially when lattice distortion can be highly dynamical in these low-dimensional ionic crystals.[34]

The shift current in $(PEA)_2PbI_4$ carried a strong multiband character. While shift current generally originates from optically excited interband transitions, the transitions do not have to occur directly between the two relevant bands. Instead, it can be mediated by intermediate states, also known as multiband processes.[35–37] Calculating the shift-current integrand in a velocity gauge, as opposed to the length gauge, allowed us to separate the two-band and multiband contributions (*Supplementary note 1*).[30,36,38] Such decomposition provided more insight about the roles that the quantum geometry and wavefunctions of the electronic bands play. In $(PEA)_2PbI_4$, the shift current was found to be largely dominated by multiband processes. Contributions from two-band transitions accounted for less than 1% of the total shift conductivity (Fig. 3e). This phenomenon can be understood with the high density of states available for multiband processes. It also explained why the spectral peak, in both the calculated and measured, appeared at 0.5 eV above the bandgap energy, where more symmetry-breaking bands that carry quantum geometry are available (Fig. 3c).

Besides the spectral profiles, agreement was also achieved between the calculated and measured polarization dependence (Fig 2c&f, red curves). As the frequency-doubling version of second-order nonlinear optical process, SHG can also be derived from the same geometrical quantities as the shift current.[6] The two-lobe patterns were reproduced in the calculated polar maps of shift current and SHG responses, respectively (*Supplementary note 2*). Such agreement provided further support to inversion symmetry breaking being the root cause of the observed nonreciprocal and nonlinear responses.

While the calculated and measured spectra agreed well above the bandgap, discrepancies are obvious at lower energy. Two sub-bandgap peaks were observed in the measured photocurrent spectrum at 2.35 eV and 2.45 eV, respectively (Fig. 3e). They coincided with the excitonic features in optical absorption spectra.[39] On the one hand, the large binding energy (>200 meV) of the layered perovskite should prohibit these excitons from generating free charge carriers through thermal dissociation, especially at cryogenic temperature (8.8 K, <1 meV by $k_BT$); on the other hand, the spontaneous photocurrent observed at the excitonic energy had a magnitude comparable to that facilitated by interband transitions (>2.52 eV). These phenomena contradict the classical picture of excitons being bound electron-hole pairs, showing again that extrinsic origins of bulk photovoltaic effect relying on dissociated free charge carriers are not applicable here. Instead, excitons here may carry a quantum-geometric character like the electronic bands and facilitate the unidirectional charge transport. These sub-bandgap features cannot be captured by the tight-binding calculations; they are many-body effects that need to be accounted for by solving the Bethe-Salpeter equation in the particle-hole channel.[40] Given the complexity of the problem,[41–43] quantitative modeling of excitonic shift current will be reported in a separate piece of work.

**Sub-bandgap responses and their temperature dependence**

The spontaneous photocurrent, excited below and above the bandgap energy, differed in temperature dependence. The magnitude of photocurrent (and the SHG), excited with 3.49 eV light (>$E_g$) decreased monotonically as the temperature rose from 5 to 90 K (**Fig. 4a**). If the magnitudes of these two quantities reflected the degrees of inversion symmetry breaking, lowering the temperature made the crystal more noncentrosymmetric. This phenomenon appeared to be consistent with earlier studies where reduced structure symmetry was observed at lower temperatures, due to stronger intermolecular interactions.[44] Meanwhile, the positive relation between the shift current and ion displacements was backed by our tight-binding calculations (*Supplementary note 4*).

On the contrary, the magnitude of excitonic photocurrent (below the bandgap energy) was enhanced at elevated temperatures (**Fig. 4b**). This phenomenon may be explained by the Floquet model, where relaxation by an isotropic heat bath is required for excitonic shift current.[8] The reduction in radiative recombination at higher temperature can diminish the polarization of excited carriers and hence enhance the excitonic shift current.[41] We are cautious, however, that factors affecting the temperature dependence of excitonic shift current can be multifaceted which require further studies on both theoretical and experimental fronts.

Our work implied that layered halide perovskites are an ideal playground to explore the connection between broken inversion symmetry and quantum geometry. The organic-inorganic hybrid composition of the perovskites which can be finely and deliberately controlled makes layered perovskites particularly unique among other low-dimensional materials. Subtle inversion symmetry breaking that may be overlooked by diffractometry in layered hybrid perovskites but can be captured by nonlinear transport and nonlinear optical effects. The significance of quantum geometric aspect in layered perovskites should also be emphasized. In our case, the multi-quantum-well structure suppressed extrinsic effects while the multiband transition enhanced the geometric ones. We hope our work can provide reference for modeling second-order nonlinear effects such as bulk photovoltaic effect and SHG in hybrid 2D perovkites, as well as inspiration of photovoltaic applications and more exotic effects.

**Methods**

**Material synthesis.** Monolithic bulk crystals of $(PEA)_2PbI_4$ were synthesized via slow-cooling crystallization from an acidic solution of hydroiodic acid (HI) and hypophosphorous acid ($H_3PO_2$). $PbI_2$ (0.4 mmol) and PEAI (0.4 mmol) were dissolved in a mixture of 0.9 mL HI and 0.1 mL $H_3PO_2$ in a 10 mL glass vial. The solution was magnetically stirred and heated in a 100 °C water bath until the precursors were fully dissolved. The vials were then transferred to a Dewar flask water bath maintained at 100 °C and was allowed to cool gradually to room temperature over approximately 72 hours. This process yielded millimeter-scale single crystals, typically in the form of thin plates or flakes. The crystals were collected from the solution by vacuum filtration for subsequent uses.

**Device fabrication.** Thin flakes of $(PEA)_2PbI_4$, few-layer graphene, and hexagonal boron nitride (hBN) were mechanically exfoliated onto $SiO_2$/Si substrates cleaned by oxygen plasma. The van-der-Waals heterostructures were assembled in a dry-nitrogen glove box using dry transfer methods.[45–48] The bottom graphene and hBN were first transferred and cleaned by AFM (Nanosurf FlexAFM). The perovskite flake was then transferred followed by top hBN and graphene. The graphene electrodes were connected to gold fingers deposited using thermal evaporation through a shadow mask. 5-nm Cr was deposited before the 20-nm Au as an adhesion layer. The AFM image (Fig. 1C) was obtained after the device was completed. In samples for SHG measurements, the few-layer graphene and gold electrodes were eliminated to simplify the fabrication.

**Optoelectrical measurements.** The photocurrent measurements were performed in a closed-cycle helium cooling cryostat stage (Advanced Research Systems DMX GMX-20). The current was measured by Keithley 4200-SCS parameter analyzer with Keithley 4200-PA Remote PreAmps. For I-V sweeping, power-, polarization-, and temperature-dependent tests, a 355 nm continuous laser (Coherent OBIS LG 355-50) was used to excite the sample. The linear polarization angle of incident light can be varied by rotating a half-waveplate (Newport 05RP12-08. For wavelength-dependent test, photocurrent spectra were measured with a laser driven light source (Energetiq EQ-77X) and Princeton Instruments HRS-500 as a monochromator. With varied center wavelength, the monochromatic light from exit slit was coupled by fiber and focused on the sample by a 20× objective lens (Mitutoyo M Plan Apo 20X) with <10 μm beam diameter.

**SHG measurements.** As for SHG measurements, the sample was mounted in a close-cycled cryostat with temperature at 5 K. An ultrafast 784 nm fiber-based femtosecond laser system with a duration of ~85 fs at a repetition rate of 80 MHz (FemtoFiber smart 780, TOPTICA Photonics AG) was used in reflection geometry with light normally incident on the sample surface. In measurements where only the incident polarization of the fundamental laser was controlled, no linear polarizer or half-wave plate was used in the detection path. The polarization-resolved measurements were achieved by rotating the half-wave plate. The second-harmonic signals were spectrally filtered from the fundamental laser using a 390-nm band-pass filter (bandwidth 10 nm) before being detected on a silicon charge-coupled device camera and finally dispersed by a diffraction grating and analyzed by a spectrometer.

## Acknowledgements

This work was supported by the National Science Foundation under award DMR-2110814. The authors thank Dr. Jiaqi Cai from Dr. Xiaodong Xu's research group for his input on dry-transfer techniques. X.W. acknowledges the Wang Start-up funding provided by the College of Arts & Sciences at Washington University. M.C. and X.W. acknowledge support from the National

Science Foundation under Grant No. 2427149 (ExpandQISE). L.D. acknowledges the financial support from National Science Foundation under award number DMR-2110706. S.C acknowledges support from JSPS KAKENHI (No. JP23H04865), MEXT, Japan. C.L. is supported by start-up funds from Florida State University and the National High Magnetic Field Laboratory. The National High Magnetic Field Laboratory is supported by the National Science Foundation through NSF/DMR-2128556 and the State of Florida. A.T.B. and X.-X. Z. acknowledge the support from the National Science Foundation under DMR-2142703. K.W. and T.T. acknowledge support from the JSPS KAKENHI (Grant Numbers 21H05233 and 23H02052), the CREST (JPMJCR24A5), JST and World Premier International Research Center Initiative (WPI), MEXT, Japan. The work made use of the FSU CMMP user facility for device fabrication and microscopic imaging.

**Author information**

**Authors and Affiliations**

**Department of Physics, Florida State University, Tallahassee, FL, USA**
Zihan Zhang, Gang Shi, Kaitai Xiao, Peng Xiong, Efstratios Manousakis, Cyprian Lewandowski & Hanwei Gao

**Department of Physics, University of Chicago, Chicago, IL, USA**
Sihan Chen

**Department of Physics, Washington University, Saint Louis, MO, USA**
Mingfeng Chen & Xi Wang

**Department of Chemical and Environmental Engineering, Yale University, New Haven, CT, USA**
Jee Yung Park

**Department of Chemical and Environmental Engineering, Yale University, New Haven, CT, USA**
Jee Yung Park

**Institute for Solid State Physics, The University of Tokyo, Kashiwa, Chiba, Japan**
Swati Chaudhary

**Institute for Solid State Physics, The University of Tokyo, Kashiwa, Chiba, Japan**
Alejandro T. Busto & Xiao-Xiao Zhang

**Research Center for Electronic and Optical Materials, National Institute for Materials Science, Tsukuba, Japan**
Kenji Watanabe

**Research Center for Materials Nanoarchitectonics, National Institute for Materials Science, Tsukuba, Japan**
Takashi Taniguchi

**FSU Quantum Initiative, Florida State University, Tallahassee, FL, USA**
Peng Xiong, Cyprian Lewandowski & Hanwei Gao

**Department of Chemistry, Emory University, Atlanta, GA, USA**
Letian Dou
**Davidson School of Chemical Engineering, Purdue University, West Lafayette, IN, USA**
Letian Dou
**Institute of Materials Science and Engineering, Washington University, St. Louis, MO, USA**
Xi Wang
**National High Magnetic Field Laboratory, Tallahassee, FL, USA**
Cyprian Lewandowski

**Contributions**

Z.Z. and S.C. (Sihan Chen) contributed equally to this work. Z.Z. and H.G. initiated the project. J.Y.K. and L.D. synthesized the (PEA)$_2$PbI$_4$ crystals. Z.Z. and G.S fabricated the devices. Z.Z, G.S and K.X configured and performed the optoelectrical measurements. M.C, A.T.B., X.-X.Z. and X.W. performed the SHG measurements. S.C. (Sihan Chen), S.C. (Swati Chaudhary), E.M. and C.L. developed the tight-binding model and performed the calculations. K.W. and T.T. provided the hBN crystals. P.X. contributed to data analysis. All authors discussed the results and wrote the paper.

**Corresponding author**

Correspondence to Hanwei Gao (hanwei.gao@fsu.edu) or Cyprian Lewandowski (clewandowski@magnet.fsu.edu).

**Ethics declarations**

**Competing interests**

The authors declare no competing interests.

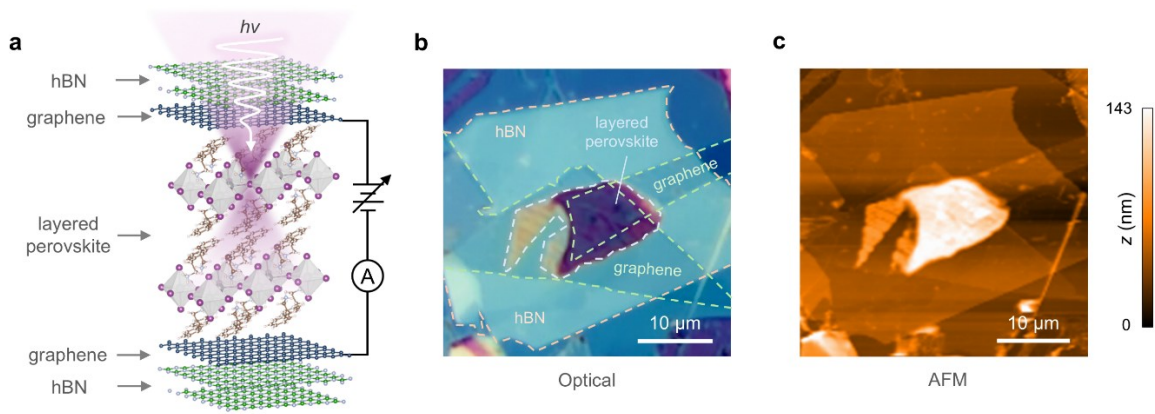

**Figure 1. Van-der-Waals heterostructures for optoelectrical measurements. (a)** Schematics of the two-terminal device with the layered perovskite sandwiched between few-layer-graphene electrodes. **(b)-(c)** Optical and atomic force microscopy images of the device. The dashed lines mark the edges of hBN, graphene, and perovskite flakes, respectively.

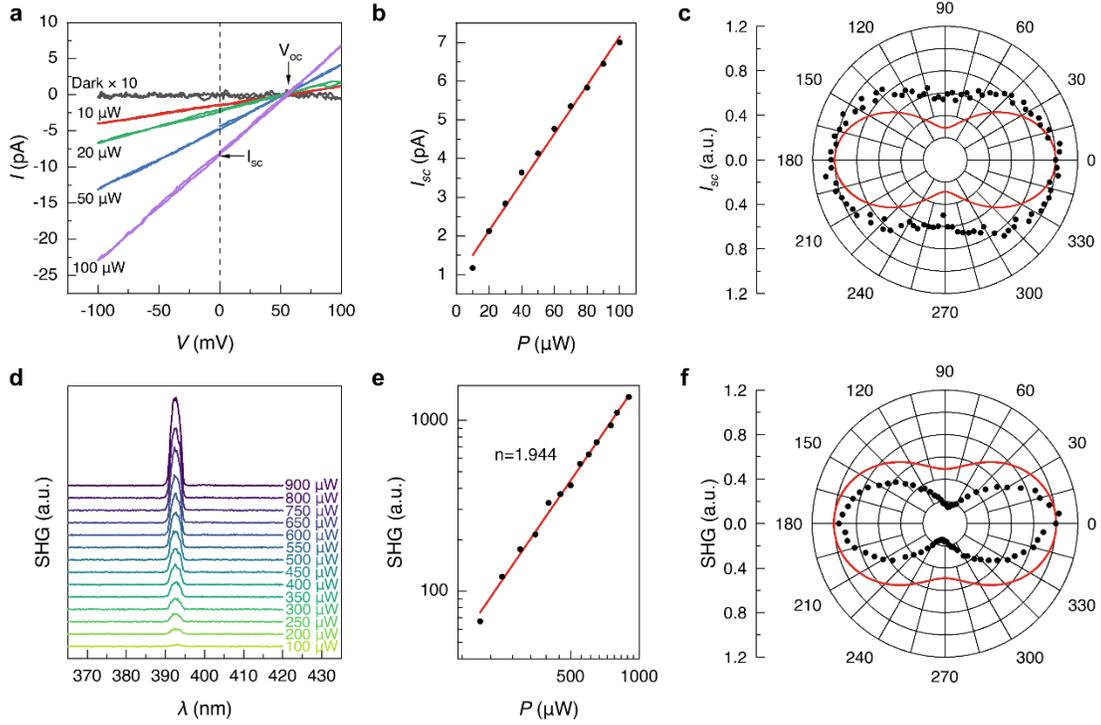

**Figure 2. Nonreciprocal and nonlinear responses of (PEA)$_2$PbI$_4$. (a)** Current-voltage characteristics in the dark and under 355-nm optical excitation, measured at a temperature of 14 K. **(b)-(c)** Dependence of the short circuit current $I_{SC}$ on the power of excitation $P$ and angle of polarization (black dots). The power dependence in (b) follows a linear function (red line). **(d)** SHG spectra of an hBN-encapsulated perovskite flake, measured at 4.5 K with varied power of optical excitation. **(e)-(f)** Dependence of the SHG intensity on the power of excitation $P$ and angle of polarization (black dots). The SHG intensity was measured as the under-the-curve area between 3.12 and 3.18 eV in the spectra. The power dependence in (e) follows a quadratic function with n ≈ 2 (red line). Red lines in (c) and (f) are the corresponding polarization dependence calculated using a tight-binding model.

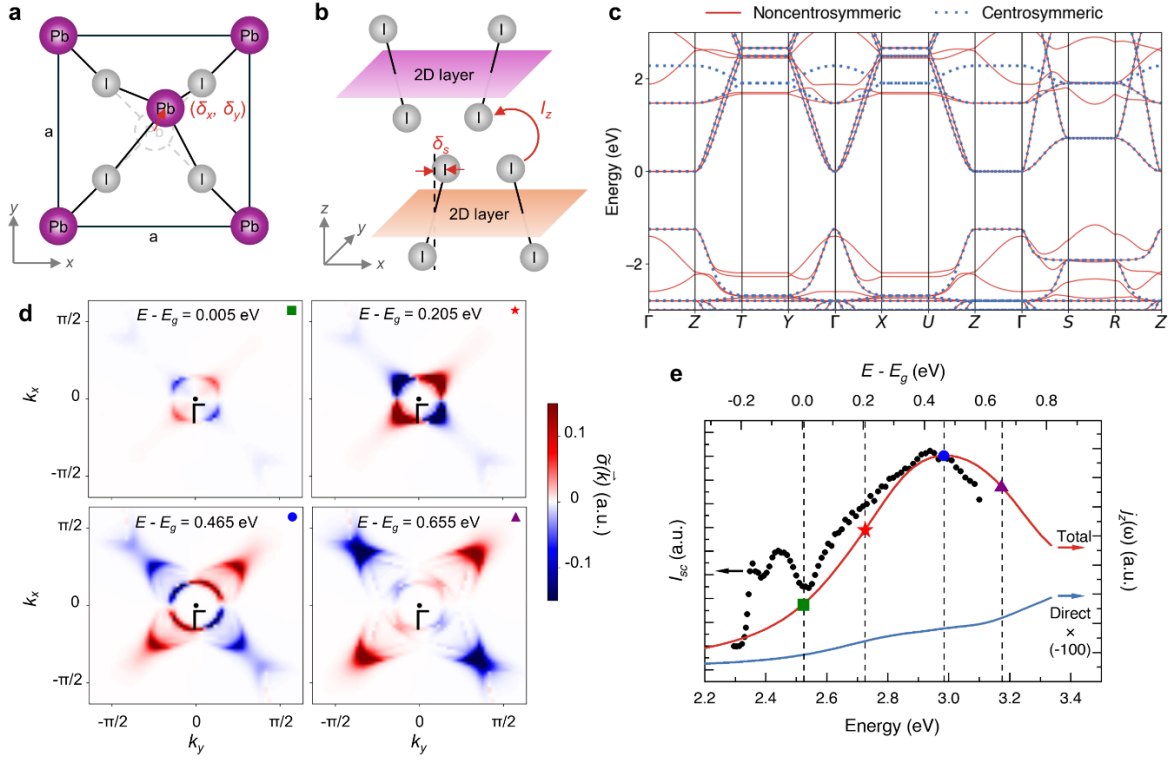

**Figure 3. An effective tight-binding model for the noncentrosymmetric crystal. (a)-(b)** Schematics of a unit cell with $Pb^{2+}$ and $I^-$ ions displaced from their centrosymmetric coordinates. An effective coupling coefficient $I_z$ was introduced to account for the electronic hopping between $I^-$ ions in adjacent layers. **(c)** The calculated electronic bands of the perovskite with (noncentrosymmetric, red solid curves) and without (centrosymmetric, blue dashed curves) the ion displacements. **(d)** Shift vector (or shift-current integrand) calculated near the $\Gamma$ point in the reciprocal space. The distribution changes drastically with the energy of interband transitions ($E$). **(e)** A comparison between the measured spontaneous photocurrent $I_{sc}$ (black dots) and calculated shift conductivity (solid curves). The contribution from direct two-band transitions (blue curve) is negligible in the calculated spectrum (red curve).

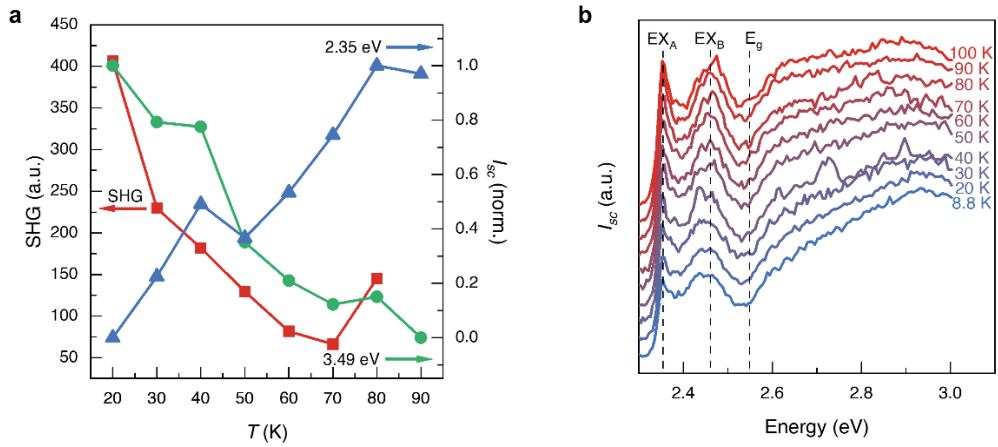

**Figure 4. Temperature dependence**. **(a)** $I_{sc}$ excited below the bandgap (excitonic) showed opposite temperature dependence compared to that above the bandgap and the SHG response. **(b)** $I_{sc}$ spectra measured at varied temperature showed enhanced excitonic features at higher temperatures.

# Supplement Information

**Supplementary Note 1: Formalism of Shift Current and SHG**

Shift current is a second-order DC effect in response to linearly polarized AC excitation.[1,2] It is characterized by a third-rank conductivity tensor $\sigma^{\beta}_{\alpha\alpha}$ and can be expressed as:

$$\mathbf{J}^{\beta} = 2\sigma^{\beta}_{\alpha\alpha}(0, -\omega, \omega)E^{\alpha}(\omega)E^{\alpha}(-\omega), \tag{1}$$

where $J^{\beta}$ is the current density, $E^{\alpha}(\omega)$ is the electric field of excitation with frequency $\omega$. Indices $\alpha$, $\beta$ denote spatial components. The shift-current conductivity is given by[1]

$$\sigma^{\beta}_{\alpha\alpha}(0, -\omega, \omega) = \frac{\pi^3}{\hbar^2} \sum_{mn} \int d^2\mathbf{k} f_{mn} \mathbf{S}^{\alpha}_{\mathbf{mn}} \left|\mathbf{A}^{\alpha}_{\mathbf{mn}}\right|^2 \delta(\omega - \epsilon_{mn}), \tag{2}$$

where $\epsilon_{mn} = \epsilon_m - \epsilon_n$ is the band energy difference between bands $n$ and $m$. The shift vector

$$\mathbf{S}^{\beta\alpha}_{mn} = \mathbf{A}^{\beta}_{mn} - \mathbf{A}^{\beta}_{nn} + \partial_{k_{\beta}} \arg\left(\mathbf{A}^{\alpha}_{mn}\right) \tag{3}$$

characterizes the real-space shift of Bloch wavefunction upon transition from band $n$ to $m$. It is also convenient to define the shift-current integrand

$$R^{\alpha\alpha\beta}_{mn} = \left|\mathbf{A}^{\alpha}_{mn}\right|^2 \mathbf{S}^{\beta\alpha}_{mn}, \tag{4}$$

which quantifies the contribution to conductivity at each point in the Brillouin zone.

Using generalized sum rules,[3] wavefunction derivatives can be replaced with sums over all states of matrix element derivatives, yielding an alternative representation of the shift-current integrand

$$R^{\alpha\alpha\beta}_{mn} = \frac{1}{\epsilon_{mn}^2} \Im \left[ \frac{h^{\alpha}_{mn} h^{\beta}_{mn} \Delta^{\alpha}_{mn}}{\epsilon_{mn}} + w^{\alpha\beta}_{mn} h^{\alpha}_{mn} \right] + \frac{1}{\epsilon_{mn}^2} \Im \left[ \sum_{n \neq m} \left( \frac{h^{\alpha}_{nm} h^{\beta}_{ml} h^{\alpha}_{ln}}{\epsilon_{ml}} - \frac{h^{\alpha}_{nm} h^{\beta}_{ml} h^{\beta}_{ln}}{\epsilon_{ln}} \right) \right], \tag{5}$$

where $h^{\alpha}_{mn} = \langle m| \partial_{\alpha} H |n\rangle$ and $w^{\alpha\beta}_{mn} = \langle m| \partial_{\alpha}\partial_{\beta} H |n\rangle$. The first term describes a direct transition between bands $n$ and $m$; the second term represents a virtual transition through intermediate bands. Depending on the density and quantum geometry of available bands, multiband processes can dominate over direct two-band transitions.[4] In our perovskite layer problem, significant contribution to conductivity only came from virtual transitions due to the large number of electronic bands.

Optical second harmonic generation (SHG) can be expressed as[5]

$$\sigma^{\beta}_{\alpha\alpha}(2\omega,\omega,\omega) = \frac{\pi e^3}{\hbar^2} \sum_{m,n} \int d^2\mathbf{k} f_{mn} S^{\beta\alpha}_{mn} |\mathbf{A}^{\alpha}_{mn}|^2 \left[ \delta(\omega - \epsilon_{mn}) - \frac{1}{2}\delta(2\omega - \epsilon_{mn}) \right]. \quad (6)$$

Apparently, SHG is also determined by the shift vector $S^{\beta\alpha}_{mn}$ and shares the same symmetry profile as the shift-current conductivity.

**Supplementary Note 2: Deriving Current Density from Conductivity**

Current density is given by[6]

$$\mathbf{J}^{\beta}(\omega \pm \omega, \omega, \omega) = \sum_{\alpha} \sigma^{\beta}_{\alpha\alpha} E^{\alpha}(\omega) E^{\alpha}(\pm\omega). \quad (7)$$

Let $E = |E|(\cos\theta, \sin\theta, 0)$, where $\theta$ is the angle between incident polarization and crystaline $x$ axis. The incident electric fields lie in the $xy$ plane and the output field travels along $z$. Under linearly polarized light, the current in the $z$-direction is

$$j^z = (\sigma^z_{xx}\cos^2\theta + \sigma^z_{yy}\sin^2\theta)|E|^2. \quad (8)$$

To compare theoretical conductivity with experiment for unpolarized light, we average over incident angles, arriving at

$$\tilde{\sigma} = \frac{1}{2}(\sigma^z_{xx} + \sigma^z_{yy}). \quad (9)$$

**Supplementary Note 3: Tight-Binding Model for the Layered Perovskite**

The method of tight-binding model was adopted from our previous work[7]. Parameters used in the model were determined by fitting with band dispersion calculated using density functional theory. In the case that ions are displaced from centrosymmetric coordinates, a single unit cell in the model contains two $Pb^{2+}$ ions and eight $I^-$ ions, each with $s$, $p_x$, $p_y$, and $p_z$-orbital degrees of freedom. Matrix elements of the TB model are given in Ref. 6. The high-symmetry paths for electronic bands follow the same convention.

*Breaking the inversion symmetry*

To break the inversion symmetry in the TB model, $Pb^{2+}$ ions were displaced from their centrosymmetric coordinates by $(\delta_x, \delta_y)$ (Fig. 3a in the main text). The effect of displacement modifies hopping coefficients by a phase term $Ve^{i\mathbf{k}(\mathbf{d}+\delta)} = V'e^{i\mathbf{k}\mathbf{d}}$.

Another mechanism of inversion symmetry breaking arises from intermolecular interactions among organic ligands docked on the inorganic layers. The resulting distortion of the octahedral framework shifts the I$^-$ ions (above and below the corresponding Pb$^{2+}$ ions) diagonally along the $xy$-plane by a small magnitude $\delta_s$ (**Fig. 3b** in the main text).

Therefore, the inversion symmetry breaking in our model was jointly determined by three structural parameters: $\delta_x$, $\delta_y$, and $\delta_s$.

*Introducing interlayer hopping*

A two-layer model was constructed by taking two copies of the single-layer tight-binding octahedral network. To facilitate transport along $z$, interlayer coupling was introduced between adjacent layers via the $p_z$ orbitals of nearest iodide ions $I^{(z)}$. In real materials, hopping between adjacent inorganic layers can be facilitated by organic ligands in-between.

When ions were displaced, each unit cell contained four pairs of interlayer-coupled iodide ions. Number the states of the $I^{(z)}$ from bottom to top as 1, 2, 3, and 4, respectively. The matrix elements can be expressed as

$$\langle 3|H|2\rangle = \langle 1|H|4\rangle = I_z e^{i\mathbf{k}\mathbf{d_3}}, \tag{10}$$

where $I_z$ is the hopping amplitude and $\mathbf{d_3}$ points from the iodide ion in the bottom layer to the top. The lateral displacement between the layers, which gave rise to the staggered stacking, was also considered.

We also considered hopping from $I^{(z)}$ $p$-orbitals to Pb $s/p$-orbitals. Denote one of the Pb$^{2+}$ ions in the bottom layer as $Pb1$ and one in the top layer as $Pb1'$. The hopping from $I^{(z)}$ to Pb$^{2+}$ is captured by the matrix elements:

$$\begin{aligned}\langle Pb1|H|1\rangle = iV_{sp}e^{i\mathbf{k}\mathbf{d_z}}, \quad &\langle 2|H|Pb1\rangle = iV_{sp}e^{i\mathbf{k}\mathbf{d_z}}, \\ \langle Pb1'|H|3\rangle = iV_{sp}e^{i\mathbf{k}\mathbf{d_z}}, \quad &\langle 4|H|Pb1'\rangle = iV_{sp}e^{i\mathbf{k}\mathbf{d_z}}.\end{aligned} \tag{11}$$

The $I^{(z)}$ ion was assumed to be directly above Pb$^{2+}$ with $\mathbf{d_z} = (0, 0, 0.1c)$. Finally, the onsite interactions of $I^{(z)}$ ions were captured by

$$\langle 1|H|1\rangle = \langle 2|H|2\rangle = \langle 3|H|3\rangle = \langle 4|H|4\rangle = E_{p_z}^{(2)}. \tag{12}$$

In the same unit cell, there are four additional $I^{(z)}$ associated with the other two Pb$^{2+}$ ions. Similarly, their matrix elements are

$$\langle 3'|H|2'\rangle = I_z e^{i\mathbf{k}\mathbf{d'_3}}, \quad \langle 1'|H|4'\rangle = I_z e^{i\mathbf{k}\mathbf{d'_3}}, \tag{13}$$

where $\mathbf{d'_3}$ points from $I^{(z)}$ in the bottom layer to the top. Their hopping and onsite interactions to corresponding Pb$^{2+}$ ions were defined similarly.

Finally, cross-hopping among the $I^{(z)}$ ions is given by

$$\langle 3'| H |2\rangle = \langle 4'| H |1\rangle = I_z e^{i\mathbf{k}\mathbf{d}_{3,c}}, \quad \langle 3| H |2'\rangle = \langle 4| H |1'\rangle = I_z e^{i\mathbf{k}\mathbf{d}'_{3,c}}. \tag{14}$$

The displacement vectors are

$$\begin{aligned} \mathbf{d_3} &= (0,\ 0.5a,\ 0.3c), & \mathbf{d'_3} &= (0,\ -0.5a,\ 0.3c), \\ \mathbf{d_{3,c}} &= (0.5a,\ 0,\ 0.3c), & \mathbf{d'_{3,c}} &= (-0.5a,\ 0,\ 0.3c), \end{aligned} \tag{15}$$

where numbers are in units of the lattice constant. Adjacent layers of the octahedral lattice were staggered, which was encoded as a 45° twist relative to each other in the theoretical model.[7] The hopping magnitude $I^{(z)}$ was set by comparing theoretical results with experiment data. The magnitude of hopping from Pb to $I^{(z)}$ and onsite interactions are adopted from Ref.6 and kept at

$$\begin{aligned} V_{sp} &= 1.2\ eV, \\ E^{(2)}_{p_z} &= -0.3\ eV, \end{aligned} \tag{16}$$

respectively.

**Supplementary Note 4: Determining Ion Displacements**

The free parameters in the TB models are $\delta_x$, $\delta_y$, $\delta_s$, and $I_z$. The optimal parameters we found are $\delta_x = -0.01$, $\delta_y = -0.02$, $\delta_s = 0.03$, and $I_z = 2.11$. Physically, $I_z$ controls interlayer hopping strength, modifies the band structure along $z$, and controls the location of conductivity peaks. The signs of conductivity are controlled by Pb displacement ($\delta_x$, $\delta_y$), see **Fig. S1**. The displacement of $I^{(z)}$ given by $\delta_s$ roughly controls the width of the conductivity peak.

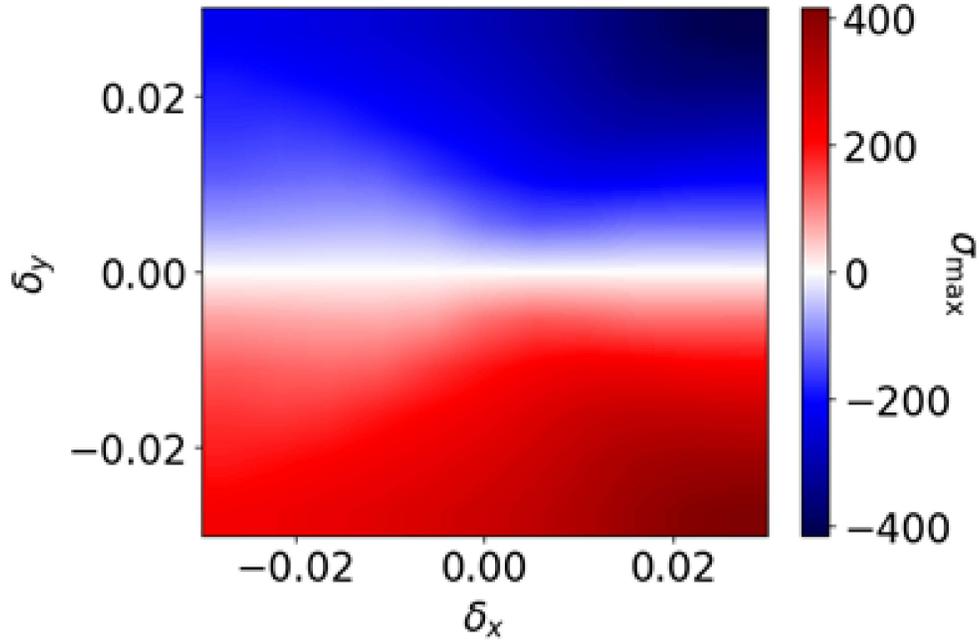

**Figure S1.** Peak conductivity for unpolarized light due to a shift-current mechanism as a function of Pb displacement. In the absence of displacement, shift current vanishes as expected.